# On non-uniqueness issues related to the permeability of a porous medium with a random porous structure

S.M. Rezaei Niya[*], S. Naghshbandi, A.P.S. Selvadurai[1][†]

[1] *Department of Civil Engineering and Applied Mechanics, McGill University, Montréal, Canada*

## Abstract

This paper discusses the issue of non-uniqueness of the permeability of a porous medium with a random structure. The permeability range for 12,000 realizations of a random porous structure is examined using a recently-developed modelling approach, which results in up to two orders of magnitude permeability variations in low porosities. These findings are compared with previous results for 13,000 realizations with a scaled-down size, and it is shown that the permeability histogram does not converge towards a narrower spectrum using larger domain sizes. The similarity between advective transport in an ensemble of porous media and a random walk phenomenon, accepted in the literature, is revisited and the inadequacy of the assumptions employed is discussed. It is shown that the conventional definition of advective transport in a porous medium, which is generally assumed to follow Hadamard's definition of well-posedness, cannot be realised. The reconstruction of a porous structure from macroscopic parameters is in itself an *ill-posed inverse problem*. To clarify this issue, explaining the *ontic* and *epistemic* identifications of a porous medium, it is discussed that while *ontic* identification cannot be employed for transport analysis, *epistemic* identification leads to non-unique solutions. It is finally suggested that a *paradigm shift* is required for better formulation of the transport characteristics of porous media.

## Keywords

Permeability, porosity, porous media, well-posed problem, non-uniqueness, post-positivism, paradigm shift, science wars

[*] Corresponding Author, seyed.niya@rmit.edu.au

[†] *William Scott Professor* and *Distinguished James McGill Professor*

## 1. Introduction

Transport phenomena in porous media has applications to a wide range of branches in engineering and the sciences (Bear 1972; Scheidegger 1974; Nield and Bejan 2006; Bear and Cheng 2010; Ichikawa and Selvadurai 2012). Having a network of tortuous pathways through a solid skeleton with an extended surface area, porous permeable structures are present both in natural and artificial materials and are of great importance to fields such as hydrogeology, geomechanics, soil mechanics, petroleum engineering, membrane materials, catalysts, porous electrodes, biophysics, and filtration.

The published literature on transport in porous media covers decades of research in many different fields, and providing a comprehensive review is beyond the scope of this article. The mere continuation of publications in this topic, even at the basic theoretical level (Ghanbarian et al. 2013; Pisani 2016; Rezaei Niya and Selvadurai 2017, 2018; Wu et al. 2019; Rao and Bai 2020; Graczyk and Matyka 2020; Armstrong et al. 2021), is an indication that the problem has not yet been clearly defined or resolved. As will be discussed in this paper, the definition of a *porous medium* may still need to be revisited. Rezaei Niya and Selvadurai (2017, 2018) have recently investigated the permeability of a general porous structure using a statistical analysis of random porous structures, and the discussions presented here are built on the findings of this research project (Rezaei Niya and Selvadurai 2017, 2018, 2019, 2021; Selvadurai et al. 2017; Selvadurai and Rezaei Niya 2020).

The permeability of a porous medium is traditionally assumed to follow Hadamard's definition of a well-posed problem with a unique and stable result, while the non-uniqueness experienced in experimental and field measurements is largely attributed to measurement errors. While it is a trivial task to construct two similar porous structures with equal porosities, tortuosities (if not defined as a fudge factor, see Ghanbarian et al. 2013), particle sizes, pore surface areas (and probably any other parameters developed for the identification of porous structures) yet significantly different permeability measures using currently available commercial modelling packages (see Figure 1 as an example), this idea of inherent non-uniqueness of permeability is controversial but merits further discussion.

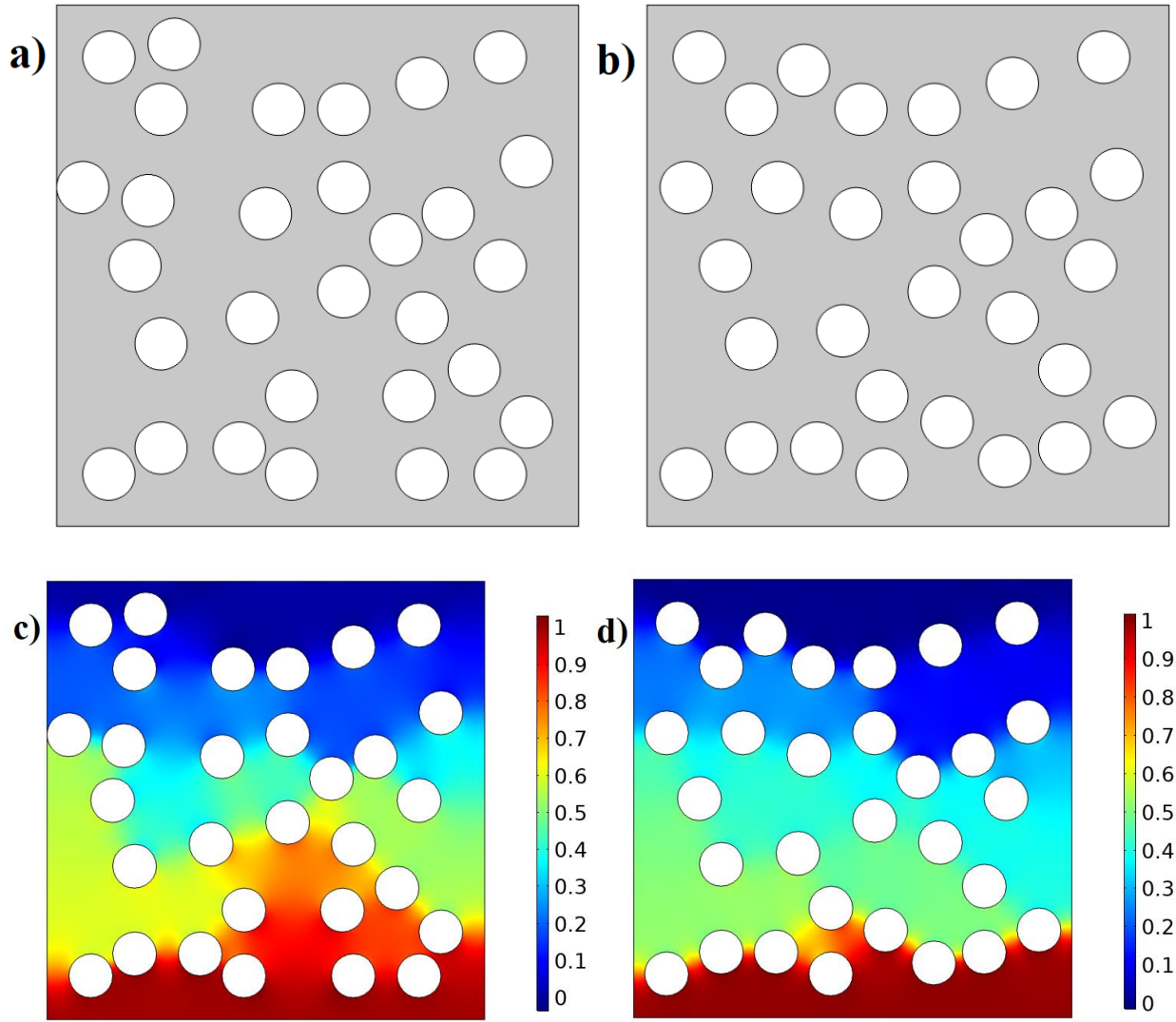

Figure 1. a,b) Two sample porous structures with equal porosities, tortuosities (Rezaei Niya and Selvadurai 2018), particle sizes, and pore surface areas, but significantly different permeabilities. The permeability of sample (a) is two times the permeability of sample (b), when the flow enters from the lower base, passes between the circular particles, and exits from the top. The side boundaries are assumed to be impermeable walls. c,d) Pressure distributions in samples (a) and (b), computed using COMSOL™ Multiphysics package.

In this manuscript, it is proposed that the permeability of a random porous structure is inherently non-unique since the advective transport problem in a porous structure is an ill-posed problem. The modelling approach employed is introduced in section 2. The modelling results that support the non-uniqueness of the permeability are presented in section 3 as permeability-porosity correlations. It is postulated that the non-uniqueness is not the result of modelling error (section 3.1) or incompetency of porosity as the indexing parameter of the porous structure (section 3.2). The literature either supporting or assuming the uniqueness of permeability is reviewed in section 4. In section 4.1, it is discussed that the non-uniqueness is not as a result of having size scales smaller than the Representative Elementary Volume (REV) and that it is also experienced at larger size scales. Section 4.2 clarifies that the non-uniqueness of the permeability of a random structure cannot be analysed when the periodicity assumption is employed in the analysis. The literature claiming fundamental similarity between a random walk process and advective transport in a porous structure is reviewed in section 4.3 and the accuracy of the assumptions employed are analysed. The well-posedness assumption of advective transport in a random porous structure is revisited in section 4.4. Section 5 addresses the fact that two fundamentally different definitions can be presented for a porous medium and non-uniqueness of the permeability is accepted when the proper definition is clarified. The need for a paradigm shift in the analysis of transport in a porous structure, and a brief discussion as to why the uniqueness of the permeability of a porous medium is generally accepted in the literature despite elementary examples to the contrary, are discussed in section 6. While, in general, permeability should be considered as a second-order tensor (Bear 1972; Ichikawa and Selvadurai 2012), it is assumed here that the randomness of the structure leads to an isotropic and homogenous porous medium at scales larger than those encountered in the identification of an REV and the permeability can be considered as a scalar. Considering the non-uniqueness of the permeability discussed in the following sections, the isotropic and homogeneous assumptions should be interpreted not as the equality of the permeability, but as the equality of statistical characteristics (e.g., average, standard deviation) of the permeability range, in different directions and positions, respectively.

## 2. Modelling approach

The analysis presented here was conducted by employing a meshless computational modelling approach recently developed by Rezaei Niya and Selvadurai (2017) for permeability analysis of porous structures. The modeling approach computes the permeability of a structure, two orders of magnitude faster than common Computational Fluid Dynamics (CFD) approaches, to within an accuracy of 10%. The network of passages in the porous structure is specified, and the ratio of the pressure drop to the flow rate for each passage is estimated. It was shown that the pressure drop of any passage can be estimated from the average length of the streamlines, which is itself calculated from the length of the surrounding boundaries of the passage. The overall flow rate and permeability of the network is then determined using a modified Hardy Cross method (Rezaei Niya and Selvadurai 2017). The 2D model developed was employed for the analysis of permeability-porosity-tortuosity correlations (Rezaei Niya and Selvadurai 2018) and was extended to special 3D cases for the analysis of flow in wormholes created by chemical erosion during acidized flow (Selvadurai et al. 2017) and flow in fractures (Rezaei Niya and Selvadurai 2019). The computational efficiency of the model has made it possible to statistically analyse the transport properties in such structures in an efficient and conventional fashion.

In the 2D model, the porous structure is defined using a hexagonal grid (Figure 2). Unlike the commonly used rectangular grid, a cell in a hexagonal grid has six neighboring cells, which significantly decreases the *disconnected area* around the cell (Rezaei Niya and Selvadurai 2017), and the developed structure more realistically resembles the physical structure of a porous medium. *Solid particles* are distributed randomly in the cells and the fluid flow between these particles is analysed. Each cell is either filled by a solid particle, or is empty and can participate in the fluid transport process. A *random* porous structure is defined here as a randomly perforated structure in which each grid is randomly specified to be either empty or contains a solid particle with a probability equal to the porosity. No type of averaging has been performed on the calculated permeabilities. The randomly-generated porous structures are analysed using the presented modelling approach and the calculated permeabilities are directly reported here. As will be discussed in section 4.2, no periodicity assumption is employed in the analysis presented here.

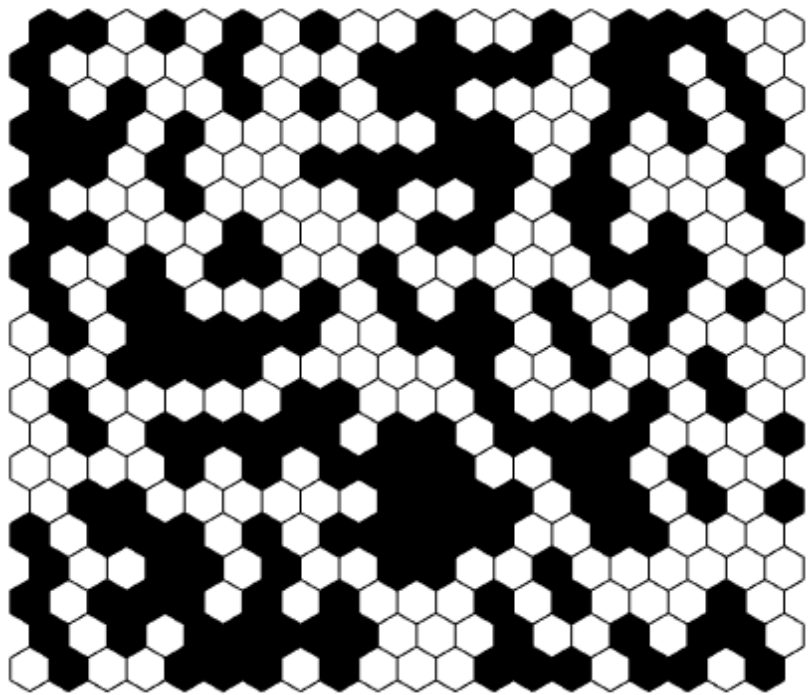

Figure 2. A sample 20x20-grid random porous structure developed for the analysis. The black cells represent *solid particles*, while the white cells participate in the fluid transport process.

## 3. Permeability-porosity correlations

The correlation between permeability and porosity, as the indexing parameter of the porous structure, has been extensively discussed in the literature (Nield and Bejan 2006; Yazdchi et al. 2011). Using a 20x20 grid, Rezaei Niya and Selvadurai (2018) recently determined the permeability of 13,000 realizations of 2D random porous structures with different porosities. These results are represented in Figure 3 along with the modelling results of a further 12,000 realizations of 2D random structures using a larger grid (40x40). The figure shows that by expanding the structure from a 20x20 grid to a 40x40 grid, there is no evidence of convergence of the results towards a unique permeability value for a specific porosity. The permeability range is even amplified at lower porosities for a 40x40 grid. The prerequisites for non-uniqueness are discussed in the following subsections.

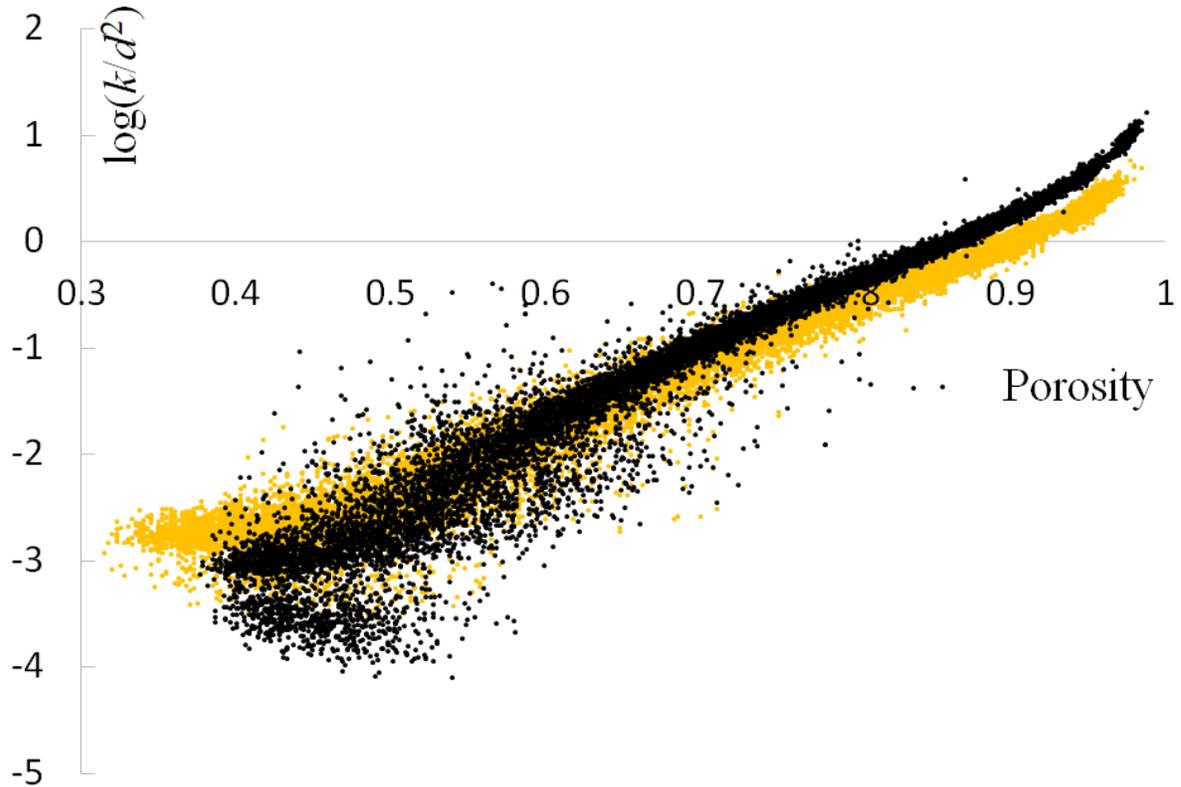

Figure 3. The calculated permeability values ($k$) for different porosities in random 2D porous structures; the darker (●) and lighter (●) dots are the results for 40x40 and 20x20 grids, respectively. Here, $d$ represents the dimension of a solid particle.

### 3.1 Reliability of the modelling results

As mentioned, the developed modelling approach has been extensively employed in different applications and its accuracy has been repeatedly verified against experimental results and numerical results obtained from commonly used fluid flow modelling software (COMSOL™ Multiphysics; Rezaei Niya and Selvadurai 2017, 2019, and ABAQUS™; Selvadurai et al. 2017) within the provided error margin. Figure 3 clearly shows that the permeability range is significantly wider than the error margin of the approach. Nevertheless, the analysis was performed on a smaller scale (due to its considerable demands on computing time) using the COMSOL™ Multiphysics Computational Fluid Dynamics (CFD) package (the results are not reported here), which also indicated that there is a significant permeability range for random structures with a specific porosity (see also Figure 1). As discussed in the following, the permeability range for random structures will be even wider (specifically at lower porosities) if this analysis is performed using commercial software.

The modelling approach explained in Section 2 and employed in this paper is prone to give a tighter permeability range than one can expect from experimental analysis of physical random structures or even numerical analysis of random structures using commercial modelling software (e.g., COMSOL[TM] Multiphysics or ABAQUS[TM]). The solid particles can only occupy the pre-defined cells in the hexagonal grid in the modelling approach employed; i.e., the horizontal spacing between the solid particles can only be a complete multiple of grid size (compare Figures 1 and 2). As a result, the diversity of the structures is significantly confined compared to a complete random distribution. In other words, the solid particles cannot be deliberately spaced/placed in this method. This constraint can significantly affect the permeability range at lower porosities. Also, the neighboring solid particles in this modelling approach are assumed to be perfectly sealed and no flow passage is possible through two neighboring particles. It is speculated that these *narrow* passages cannot significantly influence the fluid flow at higher porosities since the overall pressure drop of the structure is dominantly controlled by the *wider* passages (Rezaei Niya and Selvadurai 2018; Armstrong et al. 2021). At lower porosities, however, since the *narrow* passages are ignored, the structures that are mainly connected via *narrow* passages are specified as blocked structures and are removed from the analysis. Since these structures have significantly lower permeabilities, the permeability range of the physical porous structures is concluded to be wider than the results presented here.

It should be noted that the transport characteristics of an actual porous medium is even more complex than that discussed in this section. In this paper, it is assumed that the porous structure is fully saturated, and the capillary effects are completely ignored. Capillary effects can lead to a hysteresis due to the difference between the advancing and receding contact angles and also the ink-bottle effect (Bear 1972). Moreover, the analyses presented here are for 2D porous structures, which are substantially simpler and less complex than 3D structures (Rezaei Niya and Selvadurai 2018). In addition, it is assumed that the solid material is completely rigid and impermeable, and thereby excludes poroelastic effects (Cheng 2015; Selvadurai 2008; Selvadurai and Suvorov 2016) and any chemical/physical/thermal/mechanical interaction between the solid structure and the fluid flow can be ignored. Also, since only the low-Reynolds-number creeping flow regime of fluid flow is considered, the inertial forces and turbulent flow effects are omitted. The probable contribution from any of these parameters can widen the permeability range of a porous structure beyond the ranges discussed here.

### 3.2 Adequacy of the measure of porosity

It could be stated that the non-uniqueness of permeability of a porous structure characterized by porosity is a result of the inadequacy of the scalar porosity parameter to represent the porous structure (Bear 1972). In the literature, it is traditionally assumed that this can be solved by combining porosity with one (or more) characteristic parameter(s). Tortuosity, which has been extensively studied, is assumed to improve the permeability estimation correlations, even though there is no consensus on its precise definition (Ghanbarian et al. 2013; Rezaei Niya and Selvadurai 2018). Bear (1972) proposed another parameter, the *average medium conductance*, in his theoretical analysis, which has not been sufficiently investigated in the ensuing literature.

In a recent publication, Rezaei Niya and Selvadurai (2018) presented the correlation between permeability, porosity, tortuosity, and conductance by analysing 13,000 random porous structures. Various definitions for tortuosity and conductance were employed and the permeability-porosity-tortuosity-conductance correlations were developed. It was concluded that neither tortuosity nor conductance can improve the accuracy of the permeability-porosity correlations. A new parameter, the minimum corrected-tortuosity, was presented, which can improve the level of accuracy of the permeability estimation through correlations. The correlation between permeability and the minimum corrected-tortuosity is not one-to-one; there is still a range of permeability values for a specific minimum corrected-tortuosity, though in a tighter range, and the permeability value remains non-unique. Furthermore, a) as mentioned in Rezaei Niya and Selvadurai (2018), this parameter can only be determined during the process of permeability estimation and cannot even be calculated when the porous structure is fully known; and b) it is based on the analysis performed on a 20x20 grid and the applicability of such parameters for larger grids requires further investigation. The results presented in Section 3.3 specifically support the concept that a larger grid could be a better choice for the analysis of random structures.

Other parameters, for example particle size distribution, pore size distribution, and specific surface area, have also been employed to characterize porous structures for various applications (Bear 1972; Scheidegger 1974; Nield and Bejan 2006; Bear and Cheng 2010; Ichikawa and Selvadurai 2012). The solid particles used in the analysis here are all assumed to be of the same size; a diversity of particle sizes creates greater diversity in the possible random structures. Further analysis (not presented here) also supports the idea that combining porosity with (at least) particle size distribution will not lead to a unique permeability for a structure and variations of up to two orders of magnitude in the permeability estimates are still possible. In essence, the reconstruction of a porous structure from these parameters is an inverse problem with no unique solution, and it can be easily shown that it will not lead to a unique structure. Hence, it is reasonable to expect that such reconstruction will not result in a unique measure of permeability (see Figure 1).

The ill-posedness of the geometry reconstruction from geometrical/physical parameters is not limited to porous structures. Other complicated geometries, such as fracture surfaces (Rezaei Niya and Selvadurai 2019, 2021) and wormhole flow passages (Selvadurai et al. 2017) have similar characteristics. The fracture surface has been extensively studied and it was shown that a one-to-one correspondence between a combination of parameters and the fracture surface cannot be achieved, even when more than 20 parameters (e.g., amplitude parameters, spacing parameters, and multi-scale parameters) are employed (Smith 2014). The discussions presented in the following sections can be also employed for those geometries.

## 4. Review of literature on the question of the uniqueness of permeability

The assumption of having a unique permeability value for a random structure has been directly and indirectly justified, mostly in classical texts on transport in porous media (Bear 1972; Scheidegger 1974; Bear and Cheng 2010), although it has not been extensively discussed in recent literature. In this section, various assumptions and the analyses presented are critically reviewed and discussed.

Two points, however, need to be mentioned first: 1) This paper focuses on pressure-head-driven advective transport in porous media. It is speculated that random-walk-based transport modes (e.g., diffusion, heat conduction, electrical conduction) also result in non-unique solutions (probably with tighter deviations, Selvadurai 2019; Selvadurai and Rezaei Niya 2020), specifically close to the percolation limit (Bunde and Havlin 1991). However, this conjecture needs further detailed evaluation. 2) One can consider permeability as a *defining property* of a porous medium. The presumptions in any research conducted to date on permeability modelling and estimation did not, however, consider permeability as a defining property. Also, results similar to those presented in Figure 3 prove that there is at least a meaningful correlation between the permeability and parameters characterizing the geometry of the porous structure.

### 4.1 Representative elementary volume

It is commonly discussed in reference books on transport in porous media (Bear 1972; Ichikawa and Selvadurai 2012) that any physical (or numerical) sample of a porous medium needs to be larger than a minimum length scale to adequately define the properties of that porous medium. For smaller samples, the measured (or modelled) properties of the sample are controlled by single pores and cannot be considered as the representation of the porous structure. As a result, the Representative Elementary Volume (REV) is defined such that "the effect of adding or subtracting one or several pores has no significant influence" on its porosity (Bear 1972 p. 20; see also Selvadurai 2000 p. 5).

To estimate the REV for modelling a porous structure, 2D random structures were generated on a 1000x1000 grid with a porosity of 0.5. The porosity at selected grid sizes from the generated random structures was determined. The results for five random structures are shown in Figure 4. As seen in this figure, a 20x20 grid can be considered the lower-limit of the REV region, while larger grid sizes for transport analysis are preferred. As depicted in Figure 3, the permeability-porosity correlations for both 40x40 and 20x20 grids show similar trends. Nevertheless, the permeability range for a porosity of 0.6 is analysed using 500 random structures for each of 20x20, 40x40, and 60x60 grids; the resulted histograms are shown in Figure 5. This figure indicates that, by increasing the sample size even to 60x60 grids, there is no evidence of convergence in the histograms (as defined by moving towards narrower spectra). On the contrary, the variance of the distribution appears to increase for larger grid sizes. Interestingly, the histogram shifts from a unimodal distribution for a 20x20 grid towards a bimodal distribution for a 60x60 grid. This bimodality of the permeability distribution for low porosities can also be seen in Figure 3 (more clearly, for porosities lower than 0.5). More computationally-efficient modelling approaches and/or larger computing resources are required to extend this analysis to grids larger than 60x60. However, it can be concluded that using larger grids will not decrease the variance of the permeability distribution and certainly will not lead to a single permeability for a specific porosity.

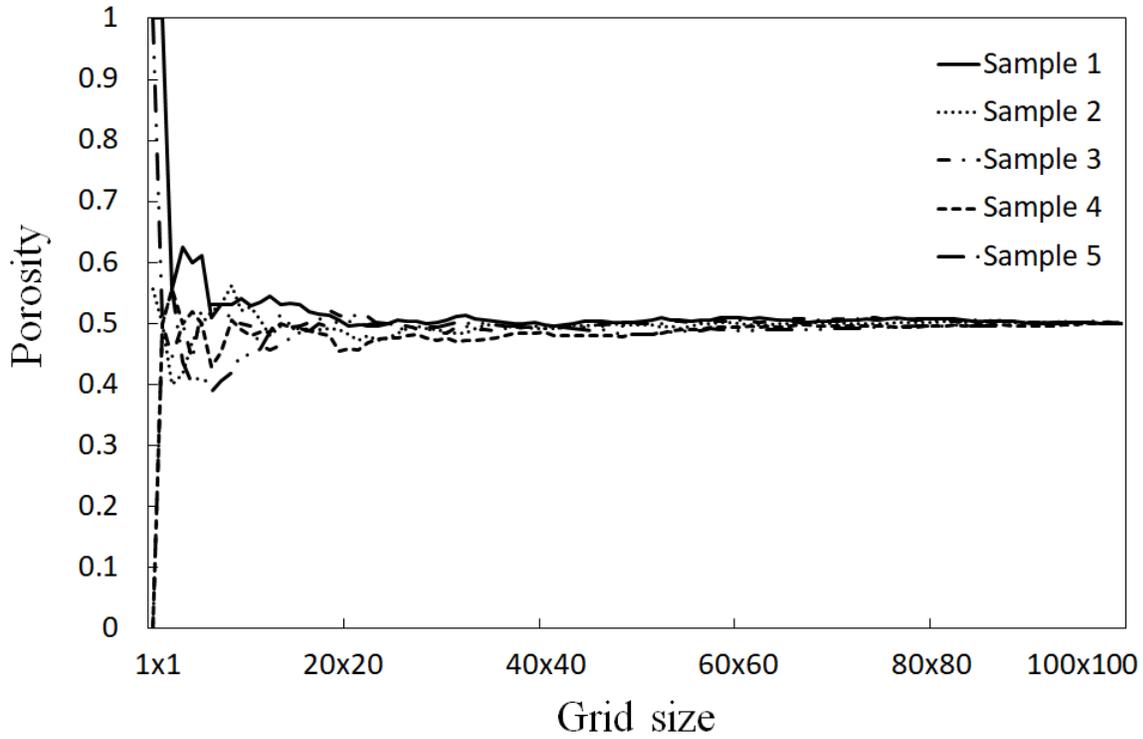

Figure 4. The porosity variations of random 2D structures with a nominal porosity of 0.5 using different grid sizes

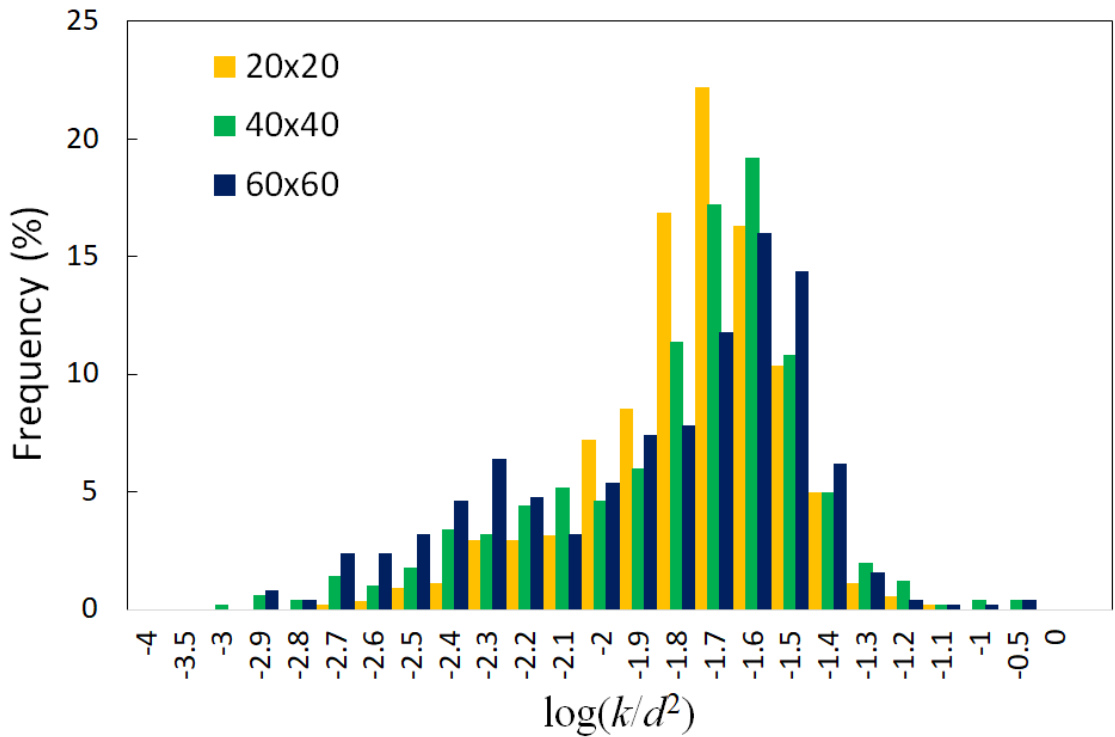

Figure 5. The histograms of permeability values of 500 random porous structures using different grid sizes for a porosity of 0.6

### 4.2 Periodicity assumption

Some analyses performed on porous structures are based on the periodicity assumption. Specifically, the volume averaging methods employ this assumption as a boundary condition (Whitaker 1999; Ichikawa and Selvadurai 2012). In these analyses, it is assumed that "the porous medium is microscopically periodic" (Ichikawa and Selvadurai 2012). With such an assumption, Darcy's law can be derived from Stokes' equation for a porous structure using a homogenization analysis (Ichikawa and Selvadurai 2012). While it is an efficient and helpful assumption, the randomness in the porous structure is not incorporated. In essence, when the periodicity assumption is employed, only a single realization of a macroscopic porous structure is analysed for each specific microscopic structure. As a result, the non-uniqueness of the permeability of a *random* structure cannot be traced in the analyses based on a periodicity assumption, while non-uniqueness of the permeability can also be experienced for periodic structures (see the Appendix).

### 4.3 Random walk conversion

The porous structures analysed in previous sections are fundamentally at smaller size scales than most of those porous media investigated in natural and artificial materials (e.g., Bear 1972; Ichikawa and Selvadurai 2012). While the results presented in Figure 5 do not support the notion of convergence towards unique permeability values at larger size scales for random structures with similar porosities, it is inherently assumed in the literature (Bear 1972; Scheidegger 1954, 1974; Bear and Cheng 2010) that such convergence is achieved at large-enough size scales, as similar convergence of a stochastic phenomenon is well documented (Rezaei Niya and Hoorfar 2016) in random-walk-based phenomena.

Some researchers have tried to convert transport in an *ensemble of porous media* to a random walk process (Dankwerts 1953; Scheidegger 1954, 1974; De Josselin de Jong 1958; Bear 1972). Random walk simulates the random or Brownian motion of particles similar to molecular diffusion in a medium, leading to the accurate modelling of diffusion-type phenomena. The random walk approach merits special attention here since it shows that the overall resultant of an infinite number of stochastic steps can lead to a unique result. If advective transport in an *ensemble of porous media* can be properly related to a random walk process, it can be concluded that the advective transport characteristics of an *ensemble of porous media* are similar to a diffusion-type process which can then lead to a unique permeability.

Scheidegger (1954, 1974) stated that the transport characteristics of an "ensemble of macroscopically identical porous media" is similar to a random walk process. He analysed the transport characteristics of an *ensemble of porous media* rather than a single porous structure and suggested that the velocity at any specific point of the ensemble can be written as the summation of an average velocity and a deviation term (Scheidegger 1954). The deviation term indicates the difference between the average velocity of the ensemble and the specific velocity of any particular porous structure. Based on the assumptions employed (specifically an ergodic assumption, discussed below), he concluded that the average of the latter term is zero and a random walk assumption for an *ensemble of porous media* is acceptable (when the ensemble is large enough). Then, due to the ergodic hypothesis, the time average can be interchanged with the ensemble average (Scheidegger 1954). Also, since different parts of one sample are considered macroscopically identical, the permeability applicable to a single porous structure is similar to the permeability of an *ensemble of porous media* and complies with the results obtained from a random walk analysis. Since Scheidegger (1954, 1974) has used ensemble averaging without volume averaging, the periodicity assumption does not directly appear in his analysis.

As emphasised by Scheidegger (1954, 1974), the ergodic hypothesis plays a crucial role in his analysis. He has stated that the ergodic hypothesis assumes the geometrical structure of any two points in the porous medium are entirely uncorrelated. Combined with isotropy and homogeneity assumptions of the porous medium, he converts the transport property for an *ensemble of porous media* elements to a random walk problem. Several points, however, need to be noted regarding these assumptions:

a) The ergodic hypothesis has been defined by Scheidegger as the non-correlation between *geometrical conditions* of any two points of the porous medium, although it has been used to imply non-correlation between *displacement* (Scheidegger 1954, 1974) and *velocity* (Bear 1972) of the fluid at any two points. However, for any specific porous medium, the fluid velocities (and displacements) of any two points are strongly correlated through mass conservation constraints. In fact, if the flow regime remains unchanged and the assumptions discussed in Section 3.1 are valid, the velocity ratio of any two points will remain unchanged, assuming they are actively involved in the transport (e.g., are not in dead-end passages). The ergodic hypothesis has then been used to estimate the average displacement after $N$ time-steps as $N$ times the average displacement of a single time-step, which is fundamentally equivalent to the periodicity assumption discussed in Section 4.2. The credibility of the ergodic hypothesis is therefore questionable, implying that interchangeability of the time average with the ensemble average over an *ensemble of porous media* is not necessarily accurate.

b) The average of the deviation term (i.e., the difference between the average velocity of the ensemble and the velocity of any specific structure) at a specific point is assumed to be zero. The randomness of the porous structure can be translated into the randomness of the hydraulic diameter of the flow passage at the specified point in any structure. While the average of the deviation term for the hydraulic diameter can be assumed to be zero, the average of the deviation term for the flow rate (and the velocity of the fluid) passing from the specified point is not necessarily zero since the flow rate is not linearly related to the hydraulic diameter. This problem arises from a subtle difference between pressure-head-driven advective transport and diffusion-type transport in a porous medium which is normally overlooked in the literature. The volumetric flux (flow rate over a unit area) of a diffusion-type process is independent of cross-sectional area (as long as the random walk assumption holds). As a result, the infinitesimal *pore* cross-sections can be integrated to estimate the flow rate of a diffusion-type process from a porous surface. The volumetric flux of an advective transport process, however, is a strong function of the cross-sectional area. As an example, if a 2D rectangular passage is divided into two parallel similar passages with half cross-sectional areas of the original passage, the diffusion-type transport rate remains the same, while the advective transport rate decreases to one quarter! As a result, the integration of cross sections of parallel passages is not an accurate evaluation of an advective transport from such structure. Therefore, the *averaging* approaches for permeability estimation based on averaging the cross section of the medium, while helpful for diffusion-type transport analysis, do not lead to accurate permeability estimations.

c) There are two assumptions that have been concluded from isotropy that need to be proved: 1) the average velocity on the ensemble at any point is aligned with the applied pressure gradient, and 2) the velocity perturbation in each direction at any point is identical.

De Josselin de Jong (1958) also studied dispersion of discrete particles in a porous medium. His assumptions, specifically, the equality of the length and conductivity of passages and the availability of passages in all directions in all the points inside the porous medium, have essentially transformed his problem description to a "Brownian motion super-imposed on a translation" (De Josselin de Jong 1958) with a residence time proportional to the direction of motion compared to the applied external pressure. The non-uniqueness of a permeability of a random porous structure cannot be expected to appear in such analysis because of the assumptions mentioned above.

The preceding discussions support the notion that transport in an *ensemble of porous media* and a *random walk process* are fundamentally dissimilar. More importantly, if a stochastic random walk process converges to a unique number, it cannot be concluded that the permeability of an ensemble of macroscopically identical porous media will be a unique value.

It should also be noted that the mere definition of the Representative Elementary Volume (REV) of a porous structure, having a porosity identical to the original medium, is inherently based on the assumption that transport characteristics of an elementary sample of the structure are accurate representations of those characteristics for the original medium, and the size scale, as long as it is larger than REV, has no noticeable effect on transport properties of a porous structure. If this assumption is considered as valid, then it should be accurate to extend non-uniqueness of permeability of a random porous structure at the REV level, shown in Figure 5, to any large-scale porous structure.

### 4.4 Well-posedness assumption

According to Hadamard (1902), a mathematical model should satisfy the basic requirements of existence, uniqueness, and stability of the solution to properly represent a physical reality. He defined such a model as a *well-posed* problem. During the last century, however, many significant and important *ill-posed* mathematical problems have been identified, which do not satisfy the well-posedness assumption (Tikhonov and Arsenin 1977; Tikhonov et al. 1995; Moritz 1995) from which inverse problems are thoroughly studied (Almasy 2007; Kabanikhin 2008). While ill-posed problems have become mathematically attractive in recent decades (Moritz 1995), the underlying assumptions of Hadamard's criteria needs specific attention. Drawing on nineteenth century understanding of a physical reality, he took it for granted that for any physical inquiry there exists a unique and stable solution. Chaotic dynamical systems have shown that a physical reality may not always lead to a stable result (Moritz 1995).

Transport in porous media, like most of other Newtonian-mechanics-based scientific advances, has drawn heavily upon nineteenth century conceptualization of physical reality in which existence, uniqueness, and stability of a solution are considered *a priori* for a valid solution. The well-posedness of the transport problem in porous media is either not discussed in the textbooks (Scheidegger 1974; Nield and Bejan 2006; Ichikawa and Selvadurai 2012) or is assumed "implicitly, without proof" (Bear 1972, p. 271; Bear and Cheng 2010, p. 204). As mentioned before, it can be easily shown that specifying a porous structure from parameters such as porosity, tortuosity, pore size or particle size distribution (or even a combination of those) is an *ill-posed* inverse problem. The problem of permeability of a porous medium, on the other hand, first needs a clear definition of what constitutes a porous medium.

## 5. What is a porous medium?

Bear (1972, p. 41) stated that,

> There is obviously no need to elaborate on the fact that a detailed description of the pore space is impossible and that, as with respect to solid grains in a granular material, only a statistical description of one kind or another is possible.

The impossibility or at least impracticality of obtaining or describing the detailed pore structure of a general porous medium has been also mentioned in other studies (Ichikawa and Selvadurai 2012). Some visualization techniques, such as X-Ray Tomography (interestingly, another ill-posed inverse problem, Moritz 1995) can help to obtain the porous structure of some specific porous media with a limited accuracy (Selvadurai et al. 2017), but the available techniques are far from perfect in visualizing the complete porous structure of a general porous medium. Nevertheless, it is assumed here that such methodology is available and applicable to all porous media and the structure can be identified and reconstructed through computer modelling.

A porous medium can be identified in two distinct ways: *Ontic identification* in which it is defined as it exists, and *epistemic identification* in which it is defined as it is perceived. A porous medium can be identified according to its unique porous structure (ontic identification). In this case, each porous medium has a unique and distinct identification. If the porous structure transforms because of poromechanical, chemical/physical interaction with the fluid, change of saturation level or permeability of the solid particles, a *new* material with a different definition emerges. In addition, no two porous media can be considered similar or categorised into one group since their identifications are fundamentally different.

On the other hand, the porous medium can be identified as it is measured or perceived (epistemic identification). In this identification, the porous medium is defined based on parameters such as porosity, tortuosity, conductance, pore distribution, or particle size distribution. As mentioned before, specifying the porous structure from such parameters is an inverse ill-posed problem with a non-unique solution. In essence, the epistemic identification presented here, is equivalent to the concept of an *ensemble of macroscopically identical porous media* introduced by Scheidegger (1954, 1974).

It needs to be emphasised that no proof is presented here that a parameter (or a finite group of parameters) cannot be defined such that it uniquely identifies a porous structure; however, the inquiries so far have not led in such a direction. As mentioned before, the published research on similar problems (e.g., characterization of surface fractures) has not been particularly promising.

The question is whether *ontic* or *epistemic* identification is implied when a porous medium is analysed. Since any theoretical analysis is, in fact, performed on a *Platonic Idealism* (here, a mathematical realization) of physical reality and characterized through a group of parameters, ontic identification cannot be intended. Furthermore, as explained previously, ontic identification cannot be obtained, at least for all the porous media at the current time. More importantly, *ontic* identification has a fundamental flaw with regards to scientific inquiry. Since a medium defined only through *ontic* identification is not characterised through (a finite group of) parameters, the results obtained for a distinct medium cannot be extended to any other medium. In logic terminology, deductive reasoning cannot be employed since no antecedent can be defined[‡].

---

[‡] Deductive or top-down reasoning draws specific conclusions from a general principle or premise. A premise is a statement from which another statement or proposition is inferred or follows as a conclusion. A primary deductive argument can be summarised as "P implies Q and P is true, therefore Q must be true." The first premise is P→Q (a conditional statement); the second premise is P (the antecedent); and the conclusion is Q (the consequent). Moritz (1995, p. 145) clarifies this as, "deduction proceeds from the general to the particular, using a general law to compute particular observable quantities."

Hawthorne (2020) says "the premises of a valid deductive argument provide total support for the conclusion". Similarly, as Moritz (1995) explains, the result of a deduction is logically correct if the process of deduction is correctly done. He calls deduction a "well-posed problem" since it is a straightforward logical process that can be precisely defined. Deductive reasoning contrasts with inductive or bottom-up reasoning, which draws general principles from specific instances (Mortiz 1995). Popper (1963) argues that induction is an ill-posed problem from which it is impossible to derive a general law.

Consequently, in any inquiry of a porous medium, the epistemic identification is intended; in other words, by a porous medium, an *ensemble of macroscopically identical porous media* is implied. While the permeability of any ontically-identified porous structure is unique, the permeability for the *ensemble* is not unique; as a result, it can be concluded that the transport phenomenon in a general porous medium has no unique solution.

It should be emphasized that the definition of *uniqueness* employed here needs specific attention. The classical linear mathematical model is considered to have a unique solution when the set of governing equations with the same boundary conditions results in one and only one solution (Selvadurai 2000). The uniqueness assumption for a physical reality, however, has two requirements: a) there is a one-to-one correspondence between any realization of the physical reality and a corresponding mathematical model; and b) the constructed mathematical model has a unique solution. Traditionally, the latter condition is violated when a physical reality is considered to have non-unique solutions (e.g., instabilities, inverse problems, Kabanikhin 2008). When there is not a one-to-one correspondence between the realizations of the physical reality and the mathematical model, the model is normally assumed to be *incomplete* and is modified (e.g., by adding more parameters to describe the physical reality). The *non-uniqueness of permeability* in the context of this paper should be interpreted as follows:

1) there is no one-to-one correspondence between permeability of a random porous structure and the current mathematical models, and

2) it is speculated that *complete* models cannot be developed due to the unlimited complexity of different realizations of this specific physical reality.

The paper does not state that successive permeability measurements of a single random structure lead to different values or does not predict that permeability measurement of a single random structure with different boundary conditions (e.g., pressures, flow rates) result in different permeability values, as long as the flow regime has not changed in the structure. In other words, permeability is unique if ontic identification is employed. It does state, however, that even if permeability of numerous porous structures is measured, permeability of a new structure cannot be accurately estimated; i.e., permeability is not unique when *epistemic* identification is employed.

Considering the discussions presented here, it can be proposed that *transport in a porous medium is a Newtonian-mechanics-based ill-posed problem with a non-unique solution*.

It could be stated here that the non-uniqueness of the solution is obtained for *random* structures, while natural (and even artificial) porous media are not in essence *random* structures. While the randomness assumption seems to be a meaningful estimation for an unknown distribution (as long as no new correlation is found), the non-uniqueness of permeability values for natural porous media have also been reported (Selvadurai et al. 2020).

## 6. Towards a paradigm shift from transport analysis in porous media

*Ontic* and *epistemic* identifications, discussed above, are in fact Natural-Scientific interpretations borrowed from philosophical perspectives in the Social Sciences. According to Webster's definition, ontology is "the science of being or reality; the branch of knowledge that investigates the nature, essential properties, and relations of being". Ontology is identified as "the starting point of all research" (Grix 2004, p. 59) that studies the various forms of being or existence. Ontological assumptions are concerned with "what constitutes reality, in other words what is" (Scotland 2012). On the other hand, *epistemology* is the theory of knowledge and how we want to know about the reality. Epistemological assumptions are concerned with "how knowledge can be created, acquired and communicated" (Scotland 2012).

Ontological and epistemological assumptions, along with methodological choices (strategies for action), and methods (specific techniques and procedures used to collect and analyse data), form a research paradigm (Crotty 1998). Moritz (1995, p. 155) notes that "a paradigm is more than a scientific theory: it is a way of thinking, a way of looking at nature". Kuhn (1962), for the first time, used the term paradigm; a paradigm is a logical system that encompasses theories, concepts, models, procedures, and techniques. Kuhn (1962) proposed a theory of scientific revolutions known as a *paradigm shift*, which implies a change in the general scientific climate. Examples of paradigm shifts in natural sciences are the acceptance of catastrophe (Guttinger and Eikemeier 1979) and chaos theories (Gleick 1987) instead of a deterministic approach.

The idea of *paradigm shift* (Kuhn 1962) offers a critique of a positivistic approach to scientific investigation, which is known as post-positivism. Ontologically, reality in the post-positivist paradigm is probabilistic and can only be known imperfectly (Katherine 2002; Robson 2002). Positivism, however, is a paradigm in the philosophy of science that emphasises observation driven by invariable natural laws and mechanisms (Fox 2008; Carpiano and Delay 2006). Hadamard's criteria of existence, uniqueness, and stability of the solution of a well-posed problem as a proper model of a physical reality, were historically and philosophically developed in a positivist paradigm, which may not necessarily be an accurate paradigm for investigation of a physical reality. The universality of the nineteenth-century positivist paradigm in scientific inquiries, sometimes fanatically defended (see the Appendix), may need a revisit for research questions with significant obstacles to further progress.

If non-uniqueness of the transport processes in random porous media and the need for a paradigm shift are accepted, the related research questions should shift from permeability-porosity correlations towards finding the statistical characteristics of the permeability of the structures with different porosities; e.g., average permeability, standard deviation, skewness, kurtosis, modality, outlier data points and their probability.

## 7. Concluding remarks

The non-uniqueness of the permeability of a random porous structure was discussed, and the published literature relevant to this issue reviewed. Using a modelling approach recently developed by the authors, the previously-published permeability values of 13,000 realizations of random porous structures were compared with 12,000 new realizations using larger grid sizes and the permeability ranges were compared. It was shown that using larger grid sizes does not lead towards a unique permeability value for random structures and the variance of the permeability distribution is even increased for larger grid sizes. The level of accuracy of the results and the representative elementary volume used for the analysis were also discussed. It was concluded that the similarity claimed between the advective transport in an ensemble of porous media and a random walk process is not convincing. It was then shown that the well-posedness assumption for transport in a random porous structure is normally assumed without any proof.

The ontic and epistemic identifications of a porous structure were explained, and it was shown that ontic identification cannot be implied when a porous structure is examined. The epistemic identification was then discussed based on an inverse ill-posed problem of reconstruction of a porous structure from statistical parameters such as porosity and particle size distribution, which leads to non-unique advective transport values for random structures. It was then suggested that analysing the transport characteristics of porous structures requires a paradigm shift towards statistical analysis of permeability histograms for various realizations of macroscopically-identical porous media.

## 8. Author contributions

The concepts related to the paper were developed by S.M.R.N. (discussions related to transport in porous media) and S.N. (conceptualization of ontological and epistemological assumptions) and verified by A.P.S.S. The paper was written by S.M.R.N. and S.N. and revised by A.P.S.S.

## 9. Competing interests

The authors declare no competing interests.

## 10. Data availability

The data that support the findings of this study are available from the corresponding author upon reasonable request.

## References

Almasy, A.A.: Inverse Problems in Classical and Quantum Physics. PhD Thesis, Universität, Mainz (2007)

Armstrong, R.T., Lanetc, Z., Mostaghimi, P., Zhuravljov, A., Herring, A., Robins, V.: Correspondence of max-flow to the absolute permeability of porous systems. Phys. Review Fluids **6**, 054003 (2021)

Bear, J.: Dynamics of Fluids in Porous Media. Dover Publications, New York (1972)

Bear, J., Cheng, A.H.-D.: Modeling Groundwater Flow and Contaminant Transport. Springer, Dordrecht (2010)

Bunde, A., Havlin, S.: Fractals and Disordered Systems. Springer-Verlag, Berlin (1991)

Carpiano, R.M., Delay, D.M.: A guide and glossary on postpositivist theory building for population health. J. Epidem. Comm. Health **60**, 564 (2006)

Cheng, A.H.D. Poroelasticity. Springer-Verlag, Berlin (2015).

Crotty, M.: The Foundations of Social Research. Sage, London (1998)

Dankwerts, P.V.: Continuous flow systems (distribution of residence times). Chem. Eng. Sci. **2**, 1 (1953)

De Josselin de Jong, G.: Longitudinal and transverse diffusion in granular deposits. Trans. Am. Geophys. Un. **39**, 67 (1958)

Fox, N.J.: Post-Positivism. In: Given, L.M. (ed.), The SAGE Encyclopedia of Qualitative Research Methods. Sage, London (2008)

Ghanbarian, B., Hunt, A.G., Ewing, P.R., Sahimi, M.: Tortuosity in porous media: a critical review. Soil Sci. Soc. Am. J. **77**, 1461 (2013)

Gleick, J.: Chaos: Making a New Science. Viking Penguin (1987)

Graczyk, K.M., Matyka, M.: Predicting porosity, permeability, and tortuosity of porous media from images by deep learning. Sci. Rep. **10**, 21488 (2020)

Grix, J.: The Foundations of Research. Macmillan Education, UK (2004)

Guttinger, W., Eikemeier, H.: Structural Stability in Physics. Springer-Verlag, Berlin (1979)

Hadamard, J. : Sur les problèmes aux dérivées partielles et leur signification physique. Princeton Univ. Bul. **49** (1902)

Hawthorne, J.: "Inductive Logic", The Stanford Encyclopedia of Philosophy. Fall 2020 Edition, Zalta, E.N. ed. (2020)

Ichikawa, Y., Selvadurai, A.P.S.: Transport Phenomena in Porous Media: Aspects of Micro/Macro Behaviour. Springer, London (2012)

Kabanikhin, S.I.: Definitions and examples of inverse and ill-posed problems. J. Inv. Ill-Posed Prob. **16**, 317 (2008)

Katherine, M.: Communication Theories: Perspectives, Processes, and Contexts. McGraw-Hill (2002)

Kuhn, T.S.: The Structure of Scientific Revolutions. Univ. Chicago Press, Chicago (1962)

Moritz, H.: Science, Mind and the Universe: An Introduction to Natural Philosophy. Heidelberg, Wichmann (1995)

Nield, D.A., Bejan, A.: Convection in Porous Media. 3rd ed., Springer, New York (2006)

Pisani, L.: A geometrical study of the tortuosity of anisotropic porous media. Transp. Porous Med. **114**, 201 (2016)

Popper, K.R.: Conjectures and Refutations: The Growth of Scientific Knowledge. Routledge, London (1963)

Rao, D., Bai, B.: Study of the factors influencing diffusive tortuosity based on pore-scale SPH simulation of granular soil. Transp. Porous Med. **132**, 333 (2020)

Rezaei Niya, S.M., Hoorfar, M.: On a possible physical origin of the constant phase element. Electrochim. Acta **188**, 98 (2016).

Rezaei Niya, S.M., Selvadurai, A.P.S.: The estimation of permeability of a porous medium with a generalized pore structure by geometry identification. Phys. Fluids **29**, 037101 (2017)

Rezaei Niya, S.M. Selvadurai, A.P.S.: A statistical correlation between permeability, porosity, tortuosity and conductance. Transp. Porous Med. **121**, 741-752 (2018)

Rezaei Niya, S.M., Selvadurai, A.P.S.: Correlation of joint roughness coefficient and permeability of a fracture. Int. J. Rock Mech. Min. Sci. **113**, 150 (2019)

Rezaei Niya, S.M., Selvadurai, A.P.S.: Modeling the approach of non-mated rock fracture surfaces under quasi-static normal load cycles. Rock Mech. Rock Eng. **54**, 1885 (2021)

Robson, C.: Real World Research: A Resource for Social Scientists and Practitioner-Researchers. Vol. 2, Oxford, Blackwell (2002)

Scheidegger, A.E.: Statistical hydrodynamics in porous media. J. Appl. Phys. **25**, 994 (1954)

Scheidegger, A.E.: The Physics of Flow Through Porous Media. 3$^{rd}$ ed., University of Toronto Press, Canada (1974)

Scotland, J.: Exploring the philosophical underpinnings of research: Relating ontology and epistemology to the methodology and methods of the scientific, interpretive, and critical research paradigms. English Lang. Teach. **5**, 9 (2012)

Selvadurai, A.P.S.: Partial Differential Equations in Mechanics. Vol. 2, Springer-Verlag, Germany (2000)

Selvadurai, A.P.S.: Interface porosity and the Dirichlet/Neumann pore fluid pressure boundary conditions in poroelasticity. Transp. Porous Med. **71**, 161 (2008)

Selvadurai, A.P.S.: A multi-phasic perspective of the intact permeability of the heterogeneous argillaceous Cobourg limestone. Sci. Rep. **9**, 17388 (2019)

Selvadurai, A.P.S., Blain-Coallier, A., Selvadurai, P.A.: Estimates for the effective permeability of intact granite obtained from the eastern and western flanks of the Canadian Shield. Minerals **10**, 667 (2020)

Selvadurai, A.P.S., Couture, C.-B., Rezaei Niya, S.M.: Permeability of wormholes created by $CO_2$-acidized water flow through stressed carbonate rocks. Phys. Fluids **29**, 096604 (2017)

Selvadurai, A.P.S., Rezaei Niya, S.M.: Effective thermal conductivity of an intact heterogeneous limestone. Rock Mech. Geotech. Eng. **12**, 682 (2020)

Selvadurai, A.P.S., Suvorov, A.P.: Coupled hydro-mechanical effects in a poro-hyperelastic material. J. Mech. Phys. Solids **91**, 311 (2016)

Smith, M.W. Roughness in the earth sciences. Earth-Sci. Rev. **136**, 202 (2014)

Tikhonov, A.N., Arsenin, V.Y.: Solutions of Ill-posed Problems. V.H. Winston & Sons, Washington (1977)

Tikhonov, A.N., Goncharsky, A.V., Stepanov, V.V., Yagola, A.G.: Numerical Methods for the Solution of Ill-posed Problems. Springer Science and Business Media, Germany (1995)

Whitaker, S.: The Method of Volume Averaging. Kluwer (1999)

Wu, H., Fang, W.Z., Kang, Q., Tao, W.Q., Qiao, R.: Predicting effective diffusivity of porous media from images by deep learning. Sci. Rep. **9**, 20387 (2019)

Yazdchi, K., Srivastava, S., Luding, S.: Microstructural effects on the permeability of periodic fibrous porous media. Int. J. Multiphase Flow **37**, 956 (2011)

**Appendix: Replies to the Reviewers**

The presented paper has been reviewed multiple times in various journals, which have resulted in extensive discussions with the reviewers (and successive rejections). While each reviewing process has turned light upon a different interpretation and aspect of the problem, and enriched the text, it has clarified for the authors that three important points, regarding the research question studied in this manuscript, are not clear for most of the reviewers:

1. It is trivial to construct porous structures with equal porosities, tortuosities (if not defined as a fudge factor! see Ghanbarian et al. 2013), particle and pore size distributions, pore surface area, and presumably any other macro parameter employed to identify the structures, yet significantly different permeability measures (similar to those shown in Figure 1). Any random porous structure, constructed in any commercial modelling package, can be easily tweaked to double or halve its permeability!

2. The classic literature *assumes* that the transport problem in a porous structure follows Hadamard's well-posedness, unanimously, without any proof, either directly (Bear 1972, p. 271; Bear and Cheng 2010, p. 204) or indirectly considering it *a priori* (Scheidegger 1974; Nield and Bejan 2006; Ichikawa and Selvadurai 2012).

3. The "non-uniqueness" of a physical parameter should always be interpreted as the impossibility of its exact prediction even when knowing exact input parameters, at least in the realm of Newtonian physics. A chaotic system, well known for its non-unique characteristics, has certainly a unique physical response, though impossible to *predict*!

The comments provided by the reviewers, along with the responses of the authors, are reported in this appendix in detail to demonstrate the reviewers' points of view and answer similar questions of the readers. A mild aggression and zealotry felt between the lines of some of the comments provided by the reviewers, presumably the distinguished experts in this field, could be interpreted as a sign that this paper stands on the borders of the current established scientific paradigm, and it therefore agitates the reviewers (most probably, unconsciously) to question the fundamental assumptions of the paradigm.

The authors of this work have together close to a century of research and publication experience and have published a couple of hundred papers before (see Prof. Selvadurai's page: https://www.mcgill.ca/civil/aps-selvadurai). This manuscript is a thoroughly interdisciplinary attempt since the philosophical discussions presented needed the contribution from a scholar with Social Science background (Dr. S. Naghshbandi).

After many discouraging discussions with the reviewers during the last year, I have learned that:

*It is a valuable contribution to the literature to write a paper that can be published; it is, however, more valuable to write a paper that cannot be published!*

S.M.R.N.

December 2021

**Replies to the <u>Reviewer #1</u>'s Comments on Physics of Fluids submission: POF21-AR-01156**

**Reviewer #1:** *I was very much interested in this work, which is aimed at analysis of permeability in a porous medium that is understood as a realization of a stochastic field. Permeability is a complex functional of the media properties and the authors' suggestion of its non-uniqueness under certain conditions is plausible. After careful reading, however, I found that this work has a few confusions or uncertainties. I focus only on major issues in my report: unless these issues are resolved and clarified, this submission is not suitable for publication.*

**Authors' Reply:** We gratefully acknowledge the comments of the reviewer. As will be discussed in detail in the following, we believe the reviewer's comments are in general the outcome of misinterpretation of the results and the conclusions which can be easily resolved with clarification of our definitions of the employed concepts. The comments have helped us to realize the potential misinterpretations of the presented discussions which have been explained in the updated version through detailed descriptions.

**Reviewer's Comment 1.** *Ensemble averaging is meaningless unless a relevant probability measure is defined. If a direct specification is difficult, the authors, at least, must fully specify the exact procedure used to generate the random porous fields they examine (the explanation on p.9 is grossly insufficient). There needs to be a sample picture of the generated domain or a typical fragment. Currently, it is impossible to make a judgment whether the proposed generation scheme is adequate.*

**Authors' Reply 1.** No ensemble averaging is performed in this work. The permeability of the randomly-generated porous structures are directly estimated numerically using the method briefly presented in section 2 (the details of the modelling approach are available in Rezaei Niya and Selvadurai, 2017). As explained in the last paragraph in page 5, the generated porous structures are 40x40 hexagonal grids where each cell is either empty with the probability equal to the porosity, or filled (blocked). A sample picture of the generated domain is now added to the manuscript (Figure 2). The estimated permeabilities of 25,000 realizations of such structures (13,000 realizations of 20x20 grids and 12,000 realizations of 40x40 grids) are presented in Figure 3. The manuscript aims at concluding the *impossibility* of accurate prediction of the permeability of a random porous structure, while this *impossibility* of prediction is translated to *non-uniqueness* of the permeability due to *impracticality* of differentiating macroscopically-similar porous structures.

The discussions related to ensemble averaging in Section 4.3 are in fact revisiting the accuracy and credibility of the analyses that previously presented in the literature to support the uniqueness of the permeability for a random porous structure.

**Reviewer's Comment 2.** *In their philosophical treatise, the authors seem to imply that the ill-posed models are necessarily inadequate (this, in fact, undermines their own work). A model is inadequate only if it is ill-posed while the underlining physical problem is well-posed. If, however, the underlining physical problem is ill-posed, then its model can (and indeed should)* *reflect this property. As I discussed below, permeability can physically be sensitive to structure under some conditions. Again, the submitted version does not provide enough information about field generation to make any specific judgement.*

**Authors' Reply 2.** The reviewer's statement about implying inadequacy of ill-posed problems in the manuscript is not correct. As clearly stated in Section 4.4, the ill-posed problems are now extensively accepted and respected in the literature. The literature, however, suggests that this has not been the case in the past (e.g., Tikhonov and Arsenin 1977; Tikhonov et al. 1995; Moritz 1995). The manuscript in fact emphasizes that permeability estimation of a random porous structure is an ill-posed physical problem which cannot be properly investigated through using well-posed models.

As explained in the authors' reply to the reviewer's first comment, the field generation process is discussed in Section 2, specifically the last paragraph.

**Reviewer's Comment 3.** *Conventional introduction of permeability in porous media is conceptually linked to volume averaging (see "The Method of Volume Averaging", 1999, Kluwer for the volume averaging theorem and its application to Darcy's law). Volume averaging can be replaced by ensemble averaging (i.e. using the ensemble averaging theorem -- see Mult. Model. Simul. 8, 1178, 2010). The authors do not seem to be familiar with standard approaches and literature. Please clearly define how is the "uniqueness of permeability" understood in this work.*

**Authors' Reply 3.** The volume averaging methods principally employ *periodicity assumption* deep in the analysis. The reference cited by the reviewer (S. Whitaker, *The Method of Volume Averaging*, Kluwer, 1999) is not an exception (see section 4.2.5, equation (4.2-21), p. 170). The reference cited by the authors in Section 4.2 has also used volume averaging to derive Darcy's law (Ichikawa and Selvadurai 2012). As discussed in Section 4.2, the non-uniqueness of the permeability of a random structure will not appear in a periodic structure since in fact only a single macroscopic porous structure is analysed when periodicity employed.

As mentioned by the reviewer, the volume averaging can be replaced by ensemble averaging, albeit with some limitations (Mult. Model. Simul. 8, 1178, 2010). The transport in a porous structure can also be analyzed using an ensemble averaging without direct employment of volume averaging and periodicity assumption (Scheidegger 1954, 1974). This method, however, is again indirectly based on periodicity assumption (see part a. in p. 14) and cannot demonstrate the non-uniqueness characteristics of the permeability.

The "standard approaches" of averaging and modelling transport in a porous structure are presented in all the classic textbooks of the field, a significant number of them cited in the manuscript (Bear 1972; Scheidegger 1974; Nield and Bejan 2006; Bear and Cheng 2010; Ichikawa and Selvadurai 2012; Selvadurai 2000, 2008).

As suggested by the reviewer, the definition of "uniqueness of permeability" in this work needs specific attention. A mathematical model is considered to have a unique solution when the set of governing equations with the same boundary conditions result in one and only one solution (e.g., Selvadurai 2000). The uniqueness assumption for a physical reality, however, has two requirements: a) there is a one-to-one correspondence between any realization of the physical reality and a corresponding mathematical model; and b) the constructed mathematical model has a unique solution. Traditionally, the latter condition is violated when a physical reality is considered to have non-unique solutions (e.g., inverse problems, Kabanikhin 2008). When there is not a one-to-one correspondence between the realizations of the physical reality and the mathematical model, the model is normally assumed *incomplete* and is modified (e.g., by adding more parameters from the physical reality). The "non-uniqueness of permeability" in the context of this manuscript should be interpreted as: 1) there is no one-to-one correspondence between permeability of a random porous structure and the so-far-presented mathematical models, and 2) it is speculated that *complete* models cannot be developed due to unlimited complexity of different realizations of this specific physical reality. The manuscript does not state that successive permeability measurements of a single random structure lead to different values, nor predicts that permeability measurement of a single random structure in different boundary conditions (e.g., pressures, flow rates) result in different permeability values, as long as the flow regime has not changed in the structure (i.e., permeability is unique if ontic identification is employed). It does state, however, that if even permeability of numerous porous structures is measured, permeability of a new structure cannot be accurately estimated (i.e., permeability is not unique if epistemic identification is employed).

**Reviewer's Comment 4.** *Irrespective of applications, the volume (or time) and ensemble averages are the same for ergodic conditions. The time or volume averages are actually measured in experiments, while ensemble averages are more often used in derivations. The authors seem to suggest that ergodicity implies the absence of correlations ("The [ergodic] hypothesis assumes that the geometrical conditions of any two points in the porous medium are entirely uncorrelated" on p.12) --- this is incorrect, and fields without any correlations are generally not a suitable model for porous media. Ergodicity requires uniformity (stationarity) and fast decay of correlations at large distances (or over long time intervals), but not the absence of correlations. To be sound, volume averaging needs to be performed over scales that are much larger than the correlation distance. It is not clear whether realizations of the medium are compliant with this constraint to produce a meaningful value of permeability.*

**Authors' Reply 4.** The authors have not claimed that the ergodic hypothesis implies the absence of correlations. This statement is in fact made by Scheidegger (1954, 1974) whose work has been criticized in Section 4.3. This point is clarified in the manuscript.

As mentioned before, any theoretical analysis based on volume averaging employs periodicity assumption as a boundary condition which then makes it out of the scope of this manuscript (as discussed in the author's reply to reviewer's comment 3). On the other hand, the Representative Elementary Volume (REV) analysis presented in Section 4.1 suggests that the numerically-generated porous structures are large enough for permeability analysis.

**Reviewer's Comment 5.** *Porous medium becomes sensitive to conditions when it is close to the percolation threshold (see "Fractals and Disordered Systems", Springer, 1991, Chpt 2). Note that pore cluster sizes can become very large near the percolation threshold increasing the size of domain needed for volume averaging. While permeability near the percolation threshold can indeed be ill-posed, there are cases that can be evaluated quite well - see the effective media approximation (Rev. Mod. Phys. 45, 574, 1973) and its generalization for random configurations of pores (Phys Rev E, 86, 011112, 2012). What is the difference between your case and the case covered by the effective media approximation?*

**Authors' Reply 5.** As mentioned by the reviewer and discussed in the provided reference (A. Bunde, S. Havlin, *Fractals and disordered systems*, Springer, 1991), the porous medium characteristics become sensitive close to the percolation threshold. However, the results presented in the manuscript in Figure 1 show that the non-uniqueness of the permeability has been experienced in all the porosity spectrum. The references cited by the reviewer are other examples of employing volume averaging approach to estimate the transport characteristics of a random porous structure. The work presented in this manuscript, however, has fundamental differences with such published literature: a) as mentioned before, no averaging has been performed in this manuscript and the calculated permeabilities of the structure are directly analyzed; b) the aim of the paper cited by the reviewer (Phys Rev E, 86, 011112, 2012) and similar literature is to *approximate* the permeability of a random porous structure. Various approximations employed in such process (e.g., Eqs. (7) and (10) in the paper cited by the reviewer) result in an average estimation rather than the whole spectrum; c) the presented paper employs a pore-network modelling approach which is not always an acceptable representation of a porous structure (see Rezaei Niya and Selvadurai 2017); and d) the assumption of similarity between diffusion-type phenomena and permeability in a porous structure is fundamentally inaccurate which will be discussed in detail in the authors' reply to the reviewer's next comment. Such inaccuracy specifically affects the underlying assumptions of averaging methods similar to the one used in the reviewer's cited paper (Phys Rev E, 86, 011112, 2012, see page 4, first paragraph in the right column).

**Reviewer's Comment 6.** *Conventional Brownian motion is a mathematically equivalent stochastic representation of diffusion (i.e. the gradient transport which can also be expressed by Om's law or by the gradient laws of creeping flows). It seems that the randomness of Brownian motion is mixed up or confused with the presumed randomness of porous media ("If transport in an ensemble of porous media can be properly related to a random walk process..." on p.11). Please clarify what is meant there. Note that the ordinary random walk can become abnormal near the percolation threshold due to trapping delays.*

**Authors' Reply 6.** The sentence aims at clarifying this point that *if* advective transport can be shown to be equivalent to a random walk process, then unique permeability for a random porous structure is probable since a random walk process, made of sufficiently large number of random steps, converges to a unique solution (e.g., diffusion or heat conduction in a medium). The sentence has been re-written to clarify this point.

It is, however, needed to be clarified that there is a subtle difference between pressure head-driven advective transport and diffusion-type transport in a porous medium which is normally overlooked in the averaging methods. The volumetric flux (flow rate over a unit area) of a diffusion-type process is independent of cross-sectional area (as long as the random walk assumption holds). As a result, the infinitesimal *pore* cross-sections can be integrated to estimate the flow rate of a diffusion-type process from a porous surface. The volumetric flux of an advective transport process, however, is a strong function of the cross-sectional area. As an example, if a 2D rectangular passage is divided to two parallel similar passages with half cross-sectional areas, the diffusion-type transport rate remains the same, while the advective transport decreases to one fourth! As a result, the integration of cross sections of parallel passages is not an accurate evaluation of an advective transport from such structure. Therefore, the *averaging* approaches for permeability estimation based on averaging the cross section of the medium, while helpful for diffusion-type transport analysis, do not lead to accurate permeability estimations.

**Reviewer's Comment 7.** *Conclusions clearly indicate what is missing in this work: a) clear definitions of volume and ensemble averaging, knowledge of relevant basic theorems, and b) strict definition of uniqueness, the scales needed for accurate volume averaging, correlation scales and possible proximity to the percolation threshold. Generally, I would not use the terms "epistemic" and "ontological" in the same way as the authors do, but I understand their thinking and believe that the authors can use any terms they prefer as long as they clearly explain their meanings.*

**Authors' Reply 7.** The reviewer's comments have been implemented into the manuscript, covering the above-mentioned points:

a) As mentioned before, no volume or ensemble averaging has been employed in this manuscript to analyze the calculated permeabilities. The volume averaging is simply out of the scope of this work since it is based on periodicity assumption. A specific type of ensemble averaging presented by Scheidegger (1954, 1974) has been criticized in detail in the manuscript.

b) The specific definition of uniqueness in this work has been explained in the updated manuscript. The difference between an advective transport and a diffusion-type transport is now discussed in more detail.

**Replies to the <u>Reviewer #2</u>'s Comments on Physics of Fluids submission: POF21-AR-01156**

**Reviewer #2:** *This paper claims that the permeability of a porous medium is a non-unique and ill-defined quantity. Aside from the method of computing the permeability that, the way the authors describe is suspect, the basis of the authors' claim is the following statement:*

*"Using a 20x20 grid, Rezaei Niya and Selvadurai [9] have recently determined the permeability of 13,000 realizations of 2D random porous structures with different porosities. These results are represented in figure 1 along with the modelling results of a further 12,000 realizations of 2D random structures using a 40x40 grid. The figure shows that by expanding the structure from a 20x20 grid to a 40x40 grid, there is no evidence of convergence of the results towards a unique permeability value for a specific porosity."*

*This argument is false. Of course, if one makes many realizations of a stochastic system, each one will have a permeability, which may be quite different from that of any other realization. That is not the key test. The test is if one AVERAGES the permeability of the realizations, and increases the number of realizations and averages again, then, provided that the size of each realization is large enough, the average will approach a constant value. If it did not, then the system does not have self-averaging property. In other words the ensemble average is not equal to the volume average.*

*The authors have not even tested whether the resolution of the grid that has been used is adequate. using only two small grids of 20 x 20 and 40 x 40 is not enough.*

*I suggest to the authors to better familiarize themselves with basic concept, and then try to "overthrow" well-established concept.*

**Authors' Reply:** The authors appreciate the reviewer's time. The reviewer's argument, however, is inaccurate in various levels:

I.    Different realizations of a stochastic system are not necessarily "quite different". Random walk theory (diffusion, heat conduction, electrical conduction, etc. in a medium) is a well-known counterexample.

II.   If a porous structure is random, it cannot be concluded that its permeability cannot be accurately determined from its porosity or other porous structure characteristics.

III.  The size scale above which "the average" approaches a constant value is called Representative Elementary Volume (REV) which is discussed in Section 4.1 of the manuscript. This size scale is not directly related to the permeability of the structure.

IV.   The reviewer has not specified whether volume averaging, ensemble averaging, time averaging, or any other type of averaging is meant in his/her comment. Each type of these averaging approaches has its own properties and limitations and should be analyzed separately.

V.    The permeability of a porous structure is by no means the average of permeabilities of its various sections. This is the core of the discussions presented in this manuscript.

VI.   Pressure head-driven advective transport is fundamentally different than a diffusion-type random-walk transport. This subtle point has been overlooked by the reviewer.

VII.  The relation between ensemble averaging and volume averaging is in general much more complicated than what is assumed by the reviewer (e.g., see I.G. Vladimirov, A.Y. Klimenko, *Multiscale Model. Simul.* 8 (2010) 1178-1211).

VIII.    The permeability histograms for grid sizes of 20x20, 40x40, and 60x60 have been analyzed and presented in Figure 5. It has been shown in Section 4.1 that a 20x20 grid can be considered as a lower-limit for REV, while 40x40 and 60x60 grids are safely larger than REV for numerical analysis.

**Replies to the <u>Editor</u>'s Comments on Physical Review Fluids submission: FE10151**

**Editor:** *The primary aim of our journal is to publish papers on novel fluid dynamics phenomena. While I am not making any judgment on the correctness or technical aspects of your work, I am afraid that it does fall outside the scope of our journal. Although your study focuses on a fluid mechanics topic (flow in porous media), I believe that there is no disagreement that inverse problems of the type you study are ill-posed, and most textbooks on the subject point this out.*

*The so-called "non-uniqueness of the permeability of a random porous structure" is well established and its discussion does not provide fundamental insight on new flow physics in my opinion.*

**Authors' Reply:** Thanks for your time reviewing our manuscript. We appreciate your feedback.

The submitted manuscript is situated in the intersection of the physical problem of transport in porous media, its mathematical interpretations, and the philosophical assumptions behind the analysis of the problem in the literature; so, we expect it to be difficult to publish due to its interdisciplinary nature. I understand it does not fit perfectly in Physical Review Fluids scope, nor any other specialized fluid mechanics or even porous media journals since the underlying philosophical assumptions are considered a priori.

As discussed in the manuscript, the well-posedness of the transport in a random porous structure is either not discussed in the textbooks (Scheidegger 1974; Nield and Bejan 2006; Ichikawa and Selvadurai 2012) or is assumed "implicitly, without proof" (Bear 1972, p. 271; Bear and Cheng 2010, p. 204). It is normally assumed that advective transport in a random porous structure is similar to a random-walk diffusion-type transport in which the outcome of sufficiently large number of stochastic steps lead to a unique result (e.g., see Scheidegger 1954, 1974; Bear 1972). That was in fact our assumption in most of our previous publications on the subject as well (Rezaei Niya and Selvadurai 2017, 2018, 2019, 2021; Selvadurai et al. 2017; Selvadurai and Rezaei Niya 2020); however, the detailed analysis of the physical problem and its mathematical literature during the last six years has changed our understanding of the problem. We tried to discuss it from our perspective that why such assumption has not been re-evaluated despite elementary counterexamples.

**Replies to the <u>Reviewer #1</u>'s Comments on Scientific Reports submission**

Submission ID 587a7b47-6194-418d-88b4-686a1fcbae41

**Reviewer #1:** *The authors discussed the so-called non-uniqueness of the permeability of porous media based on numerically predicted permeability values of a 2D random porous model. However, I cannot agree with the proposed proposition of this manuscript. As known to all, the permeability of porous media depends not only on the porosity but also other parameters including pore and particle shape and size, tortousness of capillary, pore connection etc. That is we cannot determine the permeability of porous media with porosity. Thus, it is natural that the permeability of porous media is non-unique for the porosity.*

**Authors' Reply:** It is a trivial task to reconstruct two porous media with equal porosities and parameters mentioned by the reviewer, but with significantly different permeabilities (e.g., the samples presented in Figure 1). Section 3.2 of the manuscript exclusively discusses whether a scalar porosity parameter can be accurately considered as an indexing parameter for a porous structure, and it is argued that adding other parameters will not lead towards a unique permeability measure.

**Reviewer's Comment 1.** *It is lack of up-to-date literature review in introduction part, most of the space in introduction was devoted to describe the content of this manuscript.*

**Authors' Reply 1.** While the problem is clearly a place of disagreement between the scholars of the field (and this document itself is an obvious evidence!), there is a limited recent literature on the topic, and the problem is only briefly discussed in the reference books of the field. On the other hand, Section 4 with all its sub-sections, covering more than 6 pages of the manuscript, is discussing and reviewing the literature of the topic in detail.

**Reviewer's Comment 2.** *The key method to predict the permeability of porous media should be clearly indicated rather than just show the reference.*

**Authors' Reply 2.** Section 2 of the manuscript, about one and a half page, exclusively explains the details of the method employed to estimate the permeability of porous structures.

**Reviewer's Comment 3.** *The authors have stated in section 3.1 that the proposed modeling method has been extensively verified against experimental results and numerical results obtained with commercial software. It is self-contradictory with the so-called non-uniqueness of the permeability, because the measured and numerical predicted permeability is unique for specific porous sample.*

**Authors' Reply 3.** While any method analysing any stochastic phenomenon can still be verified using statistical analysis, the proposed modelling approach has been verified against numerical results of the analysis of transport in porous media with distinct micro-level porous structures, having unique permeability measures (see Rezaei Niya and Selvadurai 2017, 2019).

**Reviewer's Comment 4.** *I think figure [3] just shows the fluctuation of the permeability for random porous media.*

**Authors' Reply 4.** Figure 3 certainly aims at showing the fluctuation of the permeability for random porous media, while emphasizing the point that completely similar random porous structures (e.g., the samples shown in Figure 1) can also exhibit significantly different permeability measures.

**Reviewer's Comment 5.** *The permeability is one of the physical properties of porous media, I cannot figure out what's the relationship between the permeability and the definition of porous media.*

**Authors' Reply 5.** Permeability has not been used to define a porous structure. It has been, however, discussed that two fundamentally different definitions can be presented for a porous medium, one of which leads to non-uniqueness of the permeability.

**Replies to the <u>Reviewer #2</u>'s Comments on Scientific Reports submission**

Submission ID 587a7b47-6194-418d-88b4-686a1fcbae41

**Reviewer #2:** *It is well known that parameter values representing sub-scale material property behavior are needed in continuum models and must be measured. If the system does not change, the estimates in permeability for a fixed system using common approaches is reliable. If the microscale structure is known, reliable estimates of permeability can be computed and this is not routinely done in many fields.*

*The premise of the manuscript is that we do not yet know of reliable estimates of permeability based upon other macroscale state quantities, such as the porosity. This is well known and do not suggest that deterministic models of transport phenomena through porous media are non-existent. It should be pointed out that permeability in natural systems is difficult to determine, sparsley measured typically, and thus flow and transport have routinely been considered stochastic for decades; many textbooks have been produced to approximate the behavior of such systems. Thus, the stochastic nature of natural porous medium systems is not in question either.*

**Authors' Reply:** The reviewer's comment, in fact, supports the discussion presented in the manuscript in a slightly different language. As mentioned by the reviewer, when "the microscale structure is known" (i.e., *ontic* identification), "reliable estimates of permeability can be computed"; however, it is not routinely done in many fields since, as discussed in the manuscript, ontic identification is not always possible. Also, "we do not yet know of reliable estimates of permeability based upon macroscale state quantities" (i.e., *epistemic* identification leads to non-uniqueness of the permeability).

The paper, however, discusses that since ontic identification (i.e., knowing the details of microscale structure) is not possible, only epistemic identification (i.e., identifying based on macroscale state quantities) is achievable and as a result, permeability is inherently non-unique. In other words, contrary to the general consensus hidden in the current literature (which can be traced between the lines in this document!), the stochastic nature experienced in measured permeabilities of natural porous media is not because of lack of experimental facilities to accurately measure the parameters such as porosity, tortuosity, particle size distribution, or pore surface area, but it is inherent in the non-unique nature of permeability of random porous structures.

**Replies to the <u>Reviewer #1</u>'s First Round of Comments on TIPM Manuscript: TIPM-D-21-00332**

**Reviewer #1:** *The authors investigate the accuracy and non-uniqueness of the calculated permeability of random porous media. The subject is clearly of interest. However, the manuscript does not bring demonstrated new facts. More, the main point, i.e. the problem of the separation of scales, is not sufficiently pointed out. My comments are listed below.*

*… In consequence, I suggest a huge revision of the paper that mainly take into account the [above] remarks.*

**Authors' Reply:** We gratefully acknowledge the comments provided by the reviewer. The reviewer's comments have helped us to significantly expand the discussions of the manuscript. The comments provided are discussed in the following detailed replies, and the changes made to the text are highlighted.

**Reviewer's Comment 1.** *The paper is supposed to address random porous structures. However figures 1 and 2 show different periodic structures. In such a case, see point 2 below.*

**Authors' Reply 1.** The clarification is as follows: Figures 1 and 2 show samples of porous structures analysed in the discussions prior to section 4.2. These structures are ***not*** samples of periodic structures of periodicity-assumption-based porous media. The reviewer's comment pointed to the fact that this aspect needs to be more clearly discussed in the manuscript; specifically, the significance of the discussions presented in section 4.3 needs to be highlighted. The manuscript has been modified accordingly and the required discussions have been appended to the text.

**Reviewer's Comment 2.** *It is of interest to first recall the case of periodic porous media. Let l be the size of the period. Then the notion of permeability K can be defined for a macroscopic sample of size L >> l at the first order of approximation of the macroscopic equivalent description. We have $K = K^{Per}(1 + O(l/L))$, where $K^{Per}$ is the permeability of a period submitted to periodic boundary conditions. Thus K is an approximative quantity , unless the porous medium is infinite. As an example, a porous medium made of 10x10x10=1000 periods will give a permeability with an approximation 10%.*

**Authors' Reply 2.** We appreciate the discussions provided by the reviewer. As mentioned before, periodic porous media have not been discussed in this manuscript since, as discussed in section 4.2, the *randomness* of the structure cannot be satisfied with the periodicity assumption. However, the point mentioned by the reviewer, that the permeability of even a periodic porous medium cannot be accurately determined, has not generally been covered in the textbooks, but is of great importance to the discussions covered in this manuscript, and has been added to the text.

**Reviewer's Comment 3.** *In the case of a random structure, an added uncertainty appears, that makes K even of poorer accuracy. And this added uncertainty is practically impossible to be* *evaluated! As recalled in the manuscript, the REV should contain a sufficiently large number of heterogeneities to overcome this added difficulty. In such a case the REV can be made periodic by introducing a thin boundary layer on its surface, with negligible consequences on the value of K (Adler and Thovert, Appl. Mech. Rev. Sep 1998, 51(9), Auriault et al, "Homogenization of coupled phenomena in heterogeneous media" 2009).*

**Authors' Reply 3.** The reviewer's comment is much appreciated. It helped us to expand the discussions of the paper, covering other aspects of the problem. The authors also specifically thank the reviewer for introducing these helpful references.

**Reviewer's Comment 4.** *A misprint on Page 8, line 3: "Section II".*

**Authors' Reply 4.** The reviewer's attention is appreciated. The misprint is revised in the updated manuscript.

**Replies to the <u>Reviewer #1</u>'s Second Round of Comments on TIPM Manuscript: TIPM-D-21-00332**

**Reviewer #1:** *The revised version does not bring sufficient answers. To develop that, it is sufficient to consider the authors' answer to my comment 1 : "The paper is supposed to address random porous structures. However figures 1 and 2 show different periodic structures." The authors respond that figures 1 and 2 do not show periodic structures. If so, what boundary conditions where applied to calculate their permeabilities? Different boundary conditions will give different permeabilities. A separation of scales is absent.*

*The added comment in section 4.2 , "... nonuniqueness of the permeability has also been reported for periodic structures (Adler and Thovert 1998, Auriault et al. 2009)", is not correct.*

*Then, unfortunately, I do not recommend the manuscript for publication.*

**Authors' Reply:** The boundary conditions have been presented in the caption of Figure 1. The same boundary conditions, classical for the analysis of non-periodic porous structures, have been employed for all the other results presented in this manuscript. Using different boundary conditions can lead to slightly different permeabilities for the same porous structures, but clearly the same boundary conditions have been employed here for all the cases analysed. The point here, however, is that having the same boundary conditions for similar porous structures result in different permeability measures (e.g., structures presented in Figure 1).

The non-uniqueness of the permeability of a periodic porous structure was in fact stated by the reviewer in previous round of comments (Comment 2). As discussed in our response to that comment, this non-uniqueness has not been reported in the literature before (to the best of our knowledge) and we therefore, cited the only references presented by the reviewer. It needs to be emphasized here that as discussed in the manuscript, only a single realization of a periodic porous structure is possible for each specific *unit cell*. As a result, *non-uniqueness* for this specific case should be interpreted as inequality of the permeability measure of the macroscopic periodic porous structure with its microscopic unit cell.

It should be mentioned that our investigation also supported the idea of possibility of non-uniqueness of the permeability measure for a periodic porous structure (although not supporting reviewer's approximation presented in Comment 2). As an example, the unit cell in Figure R1.a is vertically impermeable (black cells represent solid particles), while the periodic porous structure developed based on tiling this unit has a measurable vertical permeability (Figure R1.b). On the other hand, a periodic porous structure developed based on mirror-tiling of this unit cell (Figure R1.c) has significantly higher permeability than the unit cell (see the discussions presented in page 15, part c. in the manuscript). Obviously, the presented unit cell is a simplified and abstract example, and more elaborate structures can be developed.

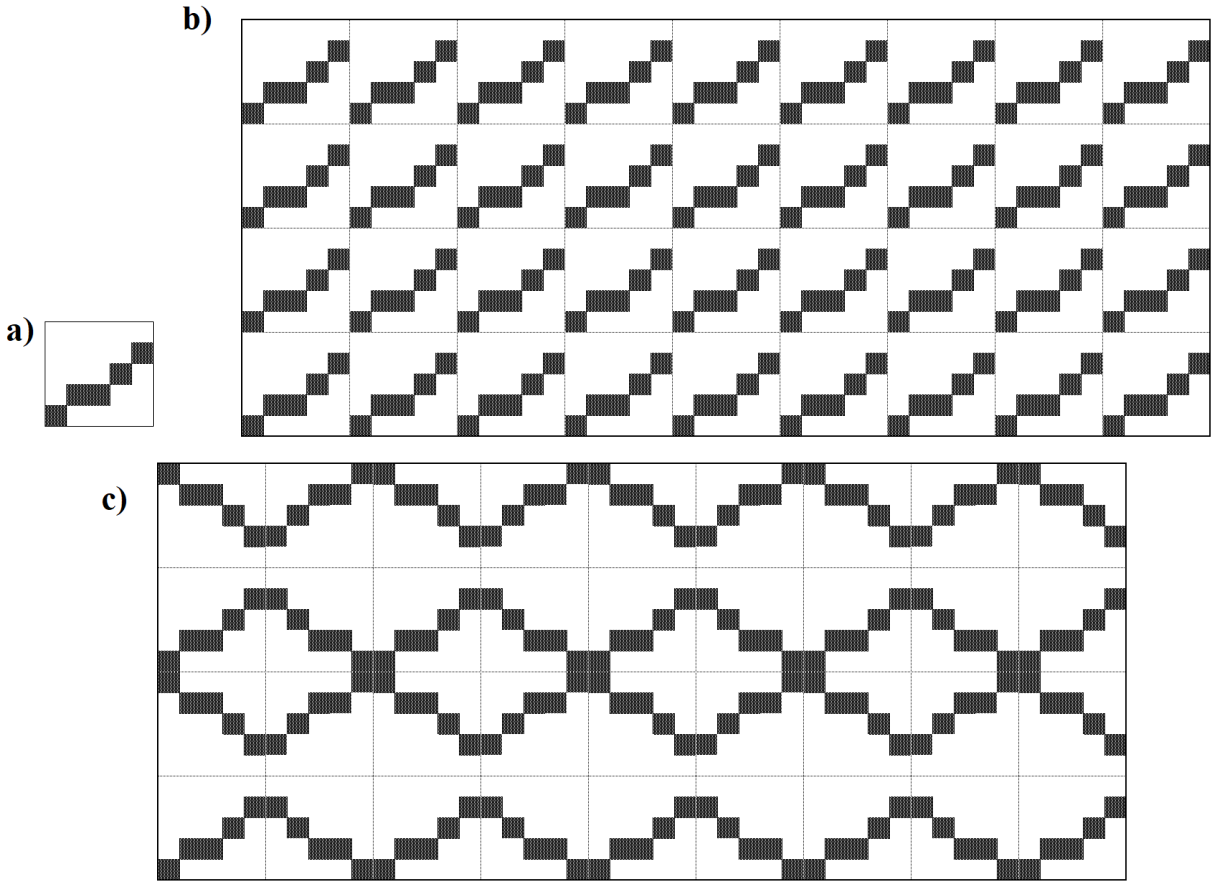

Figure R1. The possible inequality of the permeability measures of a unit cell and the periodic porous structure developed based on it. The sample unit cell (a) is vertically impermeable, while the periodic structure developed based on tiling this cell (b) has a measurable vertical permeability. On the other hand, the periodic structure developed based on mirror-tiling of the cell (c) has significantly higher permeability than the unit cell. The black cells represent solid impermeable particles.

**Replies to the <u>Reviewer #2</u>'s Comments on TIPM Manuscript: TIPM-D-21-00332**

**Reviewer #2:** *I think that this paper brings more confusion than new understanding, and that it is not suited for publication.*

*A major flaw is that the quantity under consideration, the permeability, is never defined. Thus, it is a bit preposterous to pretend to "clarify this issue" by "explaining the ontic and epistemic identifications of a porous medium" while the issue itself is obscured by lack of a proper statement. In answer to the claim that "In this manuscript, it is proposed that the permeability of a random porous structure is inherently non-unique since the advective transport problem in a porous structure is an ill-posed problem", my reply is that the question investigated in the paper is ill-formulated. Besides, it is also ill-treated.*

**Authors' Reply:** The manuscript certainly presents a non-conventional approach towards investigating fundamental philosophical assumptions considered *a priori* in this research field, and it therefore requires a meticulous revisit of these assumptions.

It is assumed here that permeability, the very basic parameter in the analysis of transport in porous structures, is well known to the readership. The reviewer is referred to the very first definitions in *any* reference book in this topic for definition of permeability, including but not limited to those referred in this manuscript (Bear 1972; Scheidegger 1974; Nield and Bejan 2006; Bear and Cheng 2010; Ichikawa and Selvadurai 2012).

The other comments provided by the reviewer are discussed in detail in the following. It will be shown that the reviewer's misunderstandings of the discussions provided in the manuscript, mainly resulted from his/her unfamiliarity with the literature of this work and his/her purely-mathematical approach to a physical reality, has resulted in inaccurate statements.

**Reviewer #2:** *The first criterion for Hadamard's well-posedness is not the unicity, but the existence of a solution. The interest of the concept of permeability is its use in upscaled, Darcy-scale flow models. In this framework, the permeability tensor should be an intrinsic coefficient relating linearly the mean fluid velocity and pressure gradient in a volume, regardless of the circumstances which induce these local features (i.e. macroscale boundary conditions). Such a tensor generally does not exist for finite samples. Different kinds of boundary conditions yield different flux/gradient relationships (for instance with no-stress or periodic conditions instead of no-flow at the side-boundaries in the example of Fig.1; by the way, I think that this example is the only case where the boundary conditions for the calculations are mentioned). This is not a matter of non-uniqueness of the permeability, but of non-existence of an intrinsic permeability tensor.*

**Authors' Reply:** The reviewer's comment clearly shows one of common misunderstandings in the literature, which has been discussed in this manuscript. Hadamard's criteria describe the *mathematical model* employed to analyse a *physical reality*. These two separate and independent entities are commonly assumed identical in the literature, while they belong to fundamentally different realms. As an example, while the criterion of the "existence of the solution" for a general mathematical model is an important question, it is meaningless here to discuss the existence of permeability for a structure. Any *physical* porous structure imagined, has a permeability, either measurable or non-accurately-measurable.

The reviewer seems to consider the porous structure only as a mathematical concept and is not familiar enough with the physics of the problem:

a) We assume by "Darcy scale" the reviewer means "large-enough scale"! Darcy's law is not about the scale of the porous structures!

b) The assumption of an "intrinsic permeability tensor" is in essence based on homogenization assumption which is meaningless in the micro-scale discussions presented in this manuscript!

c) Pressure gradient and mean fluid velocity are only "linearly" related when the flow regime inside the structure is a creeping flow (or alike).

d) It should again be emphasized that for *any* porous sample, there *certainly* is a tensor to relate pressure gradient to the flow rate as long as the flow regime has not changed inside the porous structure.

The discussions of the reviewer regarding the boundary conditions are inaccurate from different perspectives:

a) Employing periodic boundary conditions to analyse macroscale sample is meaningless. Periodicity is normally used as a boundary condition for a unit cell based on which the periodic "macroscale" porous structures are theoretically developed.

b) It is obvious that employing different boundary conditions can lead to different flux/gradient relationships!

c) No poroelastic effects have been considered here, so "no-stress" boundary condition is meaningless in this context.

**Reviewer #2:** *But then it appears that what is at stake here is not the uniqueness of the permeability of a given porous medium, but the uniqueness of the correspondence (porosity, tortuosity, particle sizes, pore surface areas, and probably other parameters ...) --> permeability. That is, a few synthetic indicators are not sufficient to infer unambiguously a permeability, whatever its definition. This is kind of obvious, and exemplified by the authors themselves for reconstructed samples by the statement (p.10) "In essence, the reconstruction of a porous structure from these parameters is an inverse problem with no unique solution, and it can be easily shown that it will not lead to a unique structure. Hence, it is reasonable to expect that such reconstruction will not result in a unique measure of permeability".*

**Authors' Reply:** As discussed in Section 4 (specifically Section 4.3) of the manuscript, there is an important body of literature analysing the similarity between transport in an ensemble of porous media and random walk process. The reviewer seems not to be familiar enough with this literature and the objective of such research, which is providing a theoretical framework for the hidden assumption of uniqueness of permeability for a large-enough porous medium. This literature has been analysed and the employed assumptions have been re-evaluated in Section 4.3 of the manuscript.

It should also be mentioned that the non-uniqueness of the correspondence is equivalent to the non-uniqueness of the permeability itself when epistemic identification is employed. The reviewer is referred to Section 5 of the manuscript.

**Reviewer #2:** *Uniqueness is ruled out. The question which remains is "what scatter can be expected, depending on the amount of synthetic information and sample size". This could be an interesting question to address, provided it is done correctly.*

**Authors' Reply:** The misunderstanding of the reviewer about "uniqueness" has already been discussed. To our knowledge, we are the only research group in the literature trying to analyse the statistical characteristics of permeability in different porosities. The reviewer is referred to our previous publications (Rezaei Niya and Selvadurai 2017, 2018, 2019, 2021; Selvadurai et al. 2017; Selvadurai and Rezaei Niya 2020).

**Reviewer #2:** *A minimal requirement would be to clearly define what is considered and to operate in clean conditions. Conditions where the arbitrary choice of boundary conditions cannot be suspected to influence the conclusions. The ideal situation is periodic media, for which the existence of an intrinsic permeability tensor is certain. It could then be examined how scattered it can be for different realizations of the unit cell with identical mean geometrical indicators, and how the mean and scatter depend on the cell size.*

**Authors' Reply:** The reviewer seems to be so faithfully engaged in the homogenization school of transport analysis in porous media that cannot see the limitations of this approach. It needs to be re-emphasized here that a) the same boundary conditions, classical for the analysis of non-periodic porous structures, have been employed for all the structures analysed in this manuscript (and all our previous publications and most of similar publications on permeability-porosity correlations cited in our publications); b) periodicity assumption is only a simplifying mathematical assumption which is essentially against the randomness in the porous structure. A periodic medium is simply not a random porous structure! Section 4.2 has been specifically added to the manuscript to clarify this issue.

**Reviewer #2:** *Unfortunately, this situation is deliberately avoided for an obscure reason (Sec. 4.2): "While [periodicity] is an efficient and helpful assumption, the randomness in the porous structure is not incorporated. In essence, when the periodicity assumption is employed, only a single realization of a macroscopic porous structure is analysed for each specific microscopic structure. As a result, the non-uniqueness of the permeability of a random structure cannot be traced in the analyses based on a periodicity assumption, while nonuniqueness of the permeability has also been reported for periodic structures (Adler and Thovert 1998, Auriault et al. 2009). "*

*Note first that the last statement is wrong. Both of these references prove that an intrinsic permeability tensor exists for a periodic medium and that it is unique. This does not preclude of course different media with different unit cells (different realizations) to have different permeabilities, even if they have identical global characteristics (porosity, surface area, ...)..*

**Authors' Reply:** The last statement was suggested (in a slightly different language) by the other reviewer and since s/he only presented the cited references, those references were reported here as well. Our investigation also supported the idea of possibility of non-uniqueness of the permeability measure for a periodic porous structure. The reviewer's statement is wrong, and the unit cell and its periodic media shown above in Figure R1 are simple counterexamples!

**Reviewer #2:** *Then, "the randomness in the porous structure" does exist, if the content of the unit cell is stochastically generated according to prescribed statistical parameters. Finally, the "macroscopic porous structure" is indeed set by the periodicity and unit cell, but without periodicity assumption, it is totally undetermined and therefore, it can certainly not be accounted for in the analysis of "the non-uniqueness of the permeability". So, there is no reason at all to rule out the only situation where a clean analysis is possible.*

**Authors' Reply:** The reviewer's deep commitment to homogenization school has again resulted in inaccurate statements. It needs to be re-emphasized that:

a) The periodicity assumption is in essence employed to avoid the randomness of a general porous structure and to simplify the transport analysis. It is still certainly possible to have randomness in the unit cell of a periodic medium, but a periodic medium cannot be considered a general random structure simply because it is periodic!

b) The homogenization method (periodicity assumption) is only one (mathematical) method of analysing porous structures, and many other (analytical, modelling, and experimental) methods and approaches have been developed in the published literature. The porous media studied in these methods are certainly not "undetermined"!

c) By "clean analysis", it seems that the reviewer means employing periodicity!

**Reviewer #2:** *I am also concerned about several methodological issues. The last one (d) is the most serious.*

a) *The boundary conditions for the calculations (and the associated definition of the "permeability") are not specified.*

**Authors' Reply:** The boundary conditions have been presented in the caption of Figure 1. The same boundary conditions, classical for the analysis of non-periodic porous structures, have been employed for all the other results presented in this manuscript. The reviewer is referred to the reference books on transport in porous media for definition of permeability.

**Reviewer #2:** *b)    About Fig.3, which supposedly " shows that by expanding the structure from a 20x20 grid to a 40x40 grid, there is no evidence of convergence of the results towards a unique permeability value for a specific porosity". Obviously, nothing can be said about convergence with only 2 values of the domain size; maybe convergence (say of the mean and dispersion) is reached with the 40x40 grids (most probably not, 40x40 is still very small, in 2d).*

**Authors' Reply:** The reviewer is referred to Figure 5 and the related discussions in the manuscript (Section 4.1), where larger grid sizes have been analysed and discussed.

**Reviewer #2:** *c)    I cannot understand why only very small samples are considered; 2d steady flow calculations are cheap, 1000x1000 samples should be a matter of seconds, and could make possible to seriously discuss the convergence of the permeability histogram when using larger domain sizes.*

**Authors' Reply:** The reviewer seems to be fundamentally unfamiliar with the governing equations of transport in porous structures and drastic computational expenses of solving those equations! The results presented in this manuscript were only made possible due to a novel estimation method developed by the authors in their previous publications (specifically, Rezaei Niya and Selvadurai 2017). The reviewer is again referred to the reference books of the field (e.g., Bear 1972; Scheidegger 1974; Nield and Bejan 2006; Bear and Cheng 2010; Ichikawa and Selvadurai 2012; Selvadurai 2000).

**Reviewer #2:** *d)       All the numerical results (or their presentation) are very perplexing. The percolation threshold for 2d honeycomb lattices is about 0.70. All the non-zero permeabilities in Fig.3 for porosities < 0.70 and all the data discussed in Sec. 4.1 (where the porosity is 0.50 or 0.60) correspond to cases where the sample is percolating nevertheless, which happens with non-zero probability because of the finite sample size. The vast majority of zero permeabilites which certainly exists is not mentioned in the figures and discussion. And the non-zero permeabilities which are treated obviously correspond to non-representative samples, which totally undermines the discussion in Sec. 4.1. As a whole, my confidence in these data, or at least in the discussion about them, is very limited.*

**Authors' Reply:** The point mentioned by the reviewer has already been discussed in detail in the manuscript (Section 3.1). The reviewer is invited to read the manuscript carefully before providing his/her comments. Unfortunately, only a limited confidence in the reviewer's evaluation is possible due to his/her unfamiliarity with the literature, inconsiderate skimming of the manuscript, and fanatic hostility to other research methods, documented in the detail in the authors' replies presented above.